\pdfoutput=1
\documentclass[aps, prl, reprint,superscriptaddress]{revtex4-1}
\input {epsf.sty}
\usepackage{graphicx,epsfig}%
\usepackage{dcolumn}
\usepackage{bm}
\usepackage{amsmath,amssymb,amsthm}
\usepackage{color} 

\begin{document}
\preprint{kkk}

\title{Directly Detecting the Edge Current in a $p_x + ip_y$ Topological Superfluid}


\author{HeeSu Byun}
\affiliation{Department of Physics, KAIST, Daejeon 34141, South Korea}

\author{Jinhoon Jeong}
\affiliation{Department of Physics, KAIST, Daejeon 34141, South Korea}
\affiliation{Korea Research Institute of Standards and Science, Daejeon 34113, South Korea}

\author{Kitak Kim}
\affiliation{Department of Physics, KAIST, Daejeon 34141, South Korea}

\author{Sang Goon Kim}
\affiliation{Korea Research Institute of Standards and Science, Daejeon 34113, South Korea}
\affiliation{Department of Physics, Chungnam National University, Daejeon 34134, South Korea}

\author{Seung-Bo Shim}
\affiliation{Korea Research Institute of Standards and Science, Daejeon 34113, South Korea}

\author{Junho Suh}
\affiliation{Korea Research Institute of Standards and Science, Daejeon 34113, South Korea}

\author{Hyoungsoon Choi}\email{h.choi@kaist.ac.kr}
\affiliation{Department of Physics, KAIST, Daejeon 34141, South Korea}


\date{\today}

\date{Version \today}

\begin{abstract}
Topological superconductors are one of the most actively studied materials these days. They are a promising candidate for hosting Majorana fermions either on their boundaries or in vortex cores. Detecting 1D edge current around the periphery of a 2D $p_x + ip_y$ superconductor would be a hallmark signature of topological superconductivity, but Majorana fermions are not amenable to electronic current measurements due to their charge neutral nature. Thermal conductivity measurements, such as thermal Hall effect, are alternatively proposed, but material synthesis must come first. Superfluid $^3$He-$A$, on the other hand, is a known $p_x + ip_y$ superfluid whose edge current can be measured with a gyroscopic technique. Here, we propose a microelectromechanical system based gyroscope that will not only have enough signal sensitivity to measure the edge current but also be used to observe dimensionality induced phase transitions between different topological superfluids.
\end{abstract}

\maketitle

\vspace{11pt}

Topological properties of Hilbert space in condensed matter systems have recently attracted wide interests in physics research \cite{has10,qi11,bee15,ban16,wit16,chi16}. A topological superconductor (TSC), in particular, a material that posseses both non-trivial topology and superconductivity/superfluidity, is actively searched \cite{bee13,and15,bee16}. It would both make our collection of topological materials more complete \cite{chi16} and serve as a platform for discovering Majorana fermions. A Majorana fermion is an exotic particle whose antiparticle is the particle itself \cite{maj37}. It can manifest itself as zero-energy quasiparticles in superconductors when the particle-hole symmetry is combined with spinless $p$-wave BCS pairing. As non-abelian quasiparticles, it also has a big implication in developing fault-tolerant quantum computing \cite{nay08}.

Simplistically, finding a TSC is equivalent to finding a $p$-wave pairing superconductors. Two of the long standing candidates for $p_x + ip_y$ chiral superconductivity are the 5/2 fractional quantum Hall state \cite{wil09} and Sr$_2$RuO$_4$ \cite{ric95,ish98,mac03}. Lack of conclusive evidence in these materials has pushed the trends towards hybrid systems of topological insulators (TI), conventional superconductors, magnetic materials, and etc \cite{fu08}. Most recently, 1D nanowire experiments present the existence of the zero-energy edge states by measuring zero-bias conduction\cite{mou12, nad14}. However, bulk order parameter of these systems are not yet identified as $p$-wave nature. Also expanding the system into higher dimensions is a non-trivial problem \cite{li16}.

In all of these activities, there is one material that have been widely overlooked, superfluid $^3$He. Superfluid $^3$He, discovered in 1971 \cite{osh72}, is the only material that is unambiguously identified as a $p$-wave pairing superfluid \cite{leg75}. There are two different superfluid phases in zero magnetic field, $^3$He-$A$ and $^3$He-$B$, which are both $p$-wave superfluids.

$^3$He-$B$ phase takes up most of the zero-field phase diagram. It is a fully gapped isotropic superfluid with $J=L+S=0$ locking the relative spin orientation with respect to the orbital orientation. It is a helical superfluid, a superfluid counterpart of the 3D TI \cite{qi09}. The bulk of the $^3$He-$B$ order parameter is fully gapped, but gapless Majorana states are expected on the surface \cite{chu09}. The existence of the surface states have been confirmed experimentally \cite{aok05,cho06} and understood in terms of the Andreev bound states \cite{and64,buc81}. Direct connection between the topology of the bulk and these surface states have not been made clear in the past. With the rise of TSCs, there is a renewed interest in these surface states \cite{mur09}. However, making an unequivocal connection between the bulk topology and the edge state is a conundrum, much like the zero-bias conduction peak being insufficient evidence of topological superconductivity in solid state systems.

For this reason, we focus our attention to the other superfluid phase, the $^3$He-$A$. Under normal circumstances, it can be found in high pressure and high temperature region. However, under magnetic field or spatial constraint, it can be extended to zero temperature at any pressure. It is a $p_x + ip_y$ Weyl superfluid with two gapless nodes along the angular momentum quantization axis. Quite recently, by measuring the mobility of electron bubbles inside the liquid $^3$He, Ikegami {\it et al.} found direct evidence of chiral and time-reversal symmetry breaking in $^3$He-$A$ \cite{ike13}. What we are proposing in this letter is to extend on such findings with a realistic experimental scheme to measure the edge current. This would mark an unambiguous confirmation of a (quasi) two dimensional $p_x + ip_y$ topological superfluid.

Consider a quasi 2D structure of $^3$He-$A$. The angular momentum quantization axis becomes perpendicular to the 2D surface and $^3$He-$A$ can be treated as an in-plane fully gapped 2D $p_x + ip_y$ chiral superfluid \cite{miz16}. Under such conditions, relevant edges are defined along the 1D periphery. Maroscopic angular momentum of $^3$He-$A$ is calculated to be $N \hbar/2 \times (\Delta/E_F )^{\gamma}$ where the value of $\gamma$ can be 0 \cite{ish77,mcc79}, 1 \cite{cro75,mer80} or 2 \cite{and61} depending on the model. The controversy is known as the angular momentum paradox \cite{sto85,kit96}.

Obviously, each Cooper pair of $^3$He atoms must carry orbital angular momentum $\hbar$. Due to a very long coherence length, $\xi \approx 100$ nm, however, there is a large overlap between Cooper pairs, and the orbital current of each Cooper pair is screened by one another. In the absence of any boundaries, this leads to cancellation of the angular momentum within the bulk of the 2D $^3$He-$A$, corresponding to the stationary ground state of the bulk. From the effective Hamiltonian of the 2D superfluid $^3$He-$A$, a topological invariant can be calculated to be non-zero as well \cite{vol88}. If specular edges are present, the screened currents all accumulate on the edge, and a chiral edge current develops corresponding to the macroscopic angular momentum of $L_{\rm edge} = N \hbar/2$ \cite{sto85,sau11,vol14,tad15,shi16}. Here $N$ is the number of $^3$He atoms. This is reminiscent of the bulk-edge correspondence in a quantum Hall system. In other words, an observation of macroscopic angular momentum of $L_{\rm edge} = N \hbar/2$ is identical to measuring the edge current in $^3$He-$A$, an outcome of the topology in bulk $^3$He-$A$. If the edges are diffusive, however, axial symmetry is broken and the angular momentum is suppressed \cite{sau11,vol14}. For this reason, it is important to have specular edges.

\begin{figure}[b]
\label{gyro}
\centerline{\epsfxsize1.0\hsize\epsffile{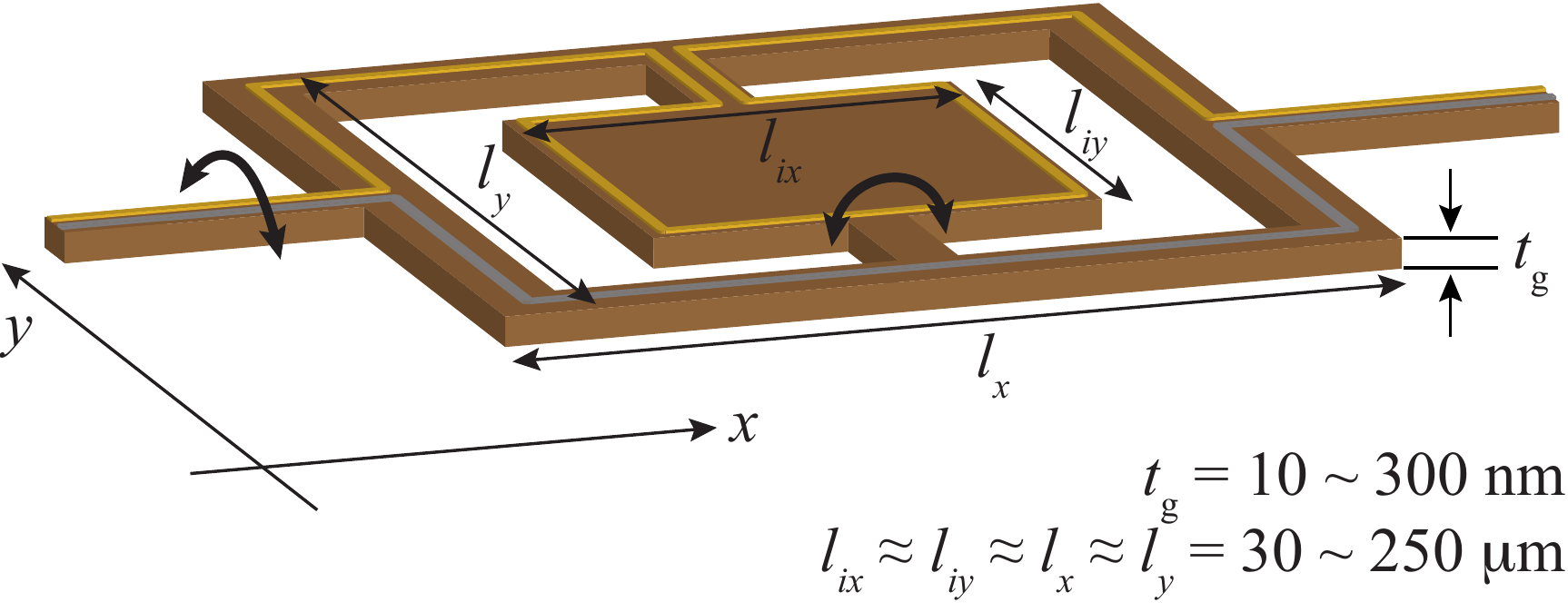}}
\begin{minipage}{1.0\hsize}
\caption {Proposed gyroscopic design not in scale. The gyroscope has two torsional modes orthogonal to each other. The device can be engineered such that the moment of inertia about the two axes are close to each other by setting $l_{ix} \approx l_{iy} \approx l_x  \approx l_y$. Our analysis in the text is carried out with such assumption. Grey and yellow lines on the gyroscope represent electrical leads for drive and detection, respectively.}
\end{minipage}
\end{figure}
\noindent

Our proposal for measuring the edge current in 2D $^3$He-$A$ is based on a micron-scale planar gyroscope as shown in Fig.~1. Without any angular momentum, a torque along the ``driving'' axis ($x$-axis), creates a rotation only about the $x$-axis. However, if there is an angular momentum about the $z$-axis in the center plate, the torque causes the angular momentum to precess, and as a result, the torque induces the center plate to rotate along the orthogonal ``sensing'' axis ($y$-axis). For a harmonic driving at the resonance frequency of torsional mode about the sensing axis, there exists a relation between the oscillation amplitude about the driving axis, $\theta_{\rm in}$, and the oscillation amplitude of the sensing axis, $\theta_{\rm out}$, given by
\begin{equation}
\label{theta_inout}
\frac{\theta_{\rm out}}{\theta_{\rm in}} = Q_y \frac{L_{\rm edge}}{I_y \omega_0}
\end{equation}
where $Q_y$, $I_y$ and $\omega_0$ is the quality factor of the resonance, moment of inertia, and resonance frequency about the sensing-axis. $L_z$ is the magnitude of the angular momentum to be detected.

In principle, the gyroscope can be applied to measuring the angular momentum $L_z = L_{\rm edge}$, but the signal has been considered undetectably small. A typical liquid helium experiment requires a container to hold the liquid of $\lesssim 1$ to 10 cm$^3$. With 10 cm$^3$ volume and a uniform macroscopic angular momentum, $L_{\rm edge} = N \hbar /2$ yields $10^{23} \hbar$. For a bulk sample, however, superfluid would nucleate in multiple regions with rotational degeneracy in picking out its angular momentum $\mathbf L$ direction. So, in reality, multiple domains of randomly oriented $\mathbf L$-vectors would cancel each other out. In addition, the $\mathbf L$-vector tends to be perpendicular to the wall of liquid helium container, and there is no avoiding the $\mathbf L$-vectors bending around the corners of a macroscopic container. In $^3$He-$A$, non-uniform $\mathbf L$-vector can also create a superflow, which would lead to spurious signal on a gyroscope \cite{vol92}.

To avoid this problem, one has to prepare a superfluid thin film whose thickness is less than the dipole length of $\sim 10~\mu$m \cite{miz16}. For reasons that will be clear later, this dimension has to be reduced down to $\sim 1~\mu$m. If the volume of the sample is reduced to 1 cm $\times$ 1 cm $\times$ 1 $\mu$m, $L_{\rm edge}$ is reduced to $10^{18} \hbar$. There has been no good way to measure angular momentum of that magnitude at ultra low temperatures. Despite its $p$-wave pairing nature, this is one of the reasons why $^3$He has not been a promising candidate for studying topological superconductivity, along with the fact that such neutral superfluid lacks charge degree of freedom.

We show in this letter that a micro-electromechanical gyroscope(MEMG) provides breakthroughs required for the measurement of the intrinsic angular momentum of $^3$He-$A$ possible. A typical device should have a nominal dimension of $\sim \rm 100~\mu m \times 100~\mu m$ with 50 to 300 nm thickness. The resonant frequency is estimated to be 10 to 100 kHz based on finite element method simulation. The measurement can be carried out by magnetomotive detection \cite{eki05}. Two sets of electrical leads, excitation and detection, can be patterned onto the device as shown in Fig.~1. In plane magnetic field along the $y$-axis, $B_y$, coupled to an ac current along the drive lead, $i(t)$, will induce an oscillatory torque $\tau_x = B_y l_x^2 i /2$ about the $x$-axis. Here $l_{x}$ is the length of the device along the $x$-axis. For the case of zero angular momentum, change in magnetic flux within the detection lead induces voltage
\begin{equation}
V = \frac{B_y^2 l_x^4}{2 I_x \omega_0} Q_x i
\label{magnetomotive}
\end{equation}
where $I_{x}$ and $Q_{x}$ represent gyroscope's moment of inertia and quality factor of the torsional oscillation about $x$-axis, respectively. Non-zero angular momentum would induce angular displacement on the orthogonal axis $\theta_{\rm out}$. With application of another magnetic field $B_x$, the voltage from this orthogonal oscillation can be detected
\begin{eqnarray}
\label{magnetomotive}
V & = & \frac{B_x B_y l_x^2 l_y^2}{2I_x \omega_0} Q_x i \times \frac{\theta_{\rm out}}{\theta_{\rm in}} \\
& = & \frac{B_x B_y l_x^2 l_y^2 L_{\rm edge}}{2 I_x I_y \omega_0^2} Q_x Q_y i \approx 100 \frac{B_x B_y n \hbar t_{\rm f}}{\alpha^2 \rho G t_{\rm g}^2} Q_x Q_y i. \nonumber
\end{eqnarray}
The frequencies of the oscillation about the two orthogonal axes, $x$ and $y$, are considered to be perfectly matched at $\omega_0$ in this case. The angular momentum is given by $L_{\rm edge} = n l_x l_y t_{\rm f} \hbar /2$ where $n \approx 2 \times 10^{22}$ cm$^{-3}$ is the number density of $^3$He, $l_{x(y)}$ is the length of the device along the $x(y)$-axis. Superfluid film thickness is given by $t_{\rm f}$. The moment of inertia $I_{x(y)}$ about $x(y)$ axis is given by $ I_{x(y)} = \approx \rho l_x l_y t_{\rm g} l_{x(y)} ^2 /12 $ where $\rho$ is the mass density of the gyroscope material and $t_{\rm g}$ is the thickness of the gyroscope membrane. The condition $l_{x(y)} \gg t_g$ is used. The resonant frequency $\omega_0$ also depends on the size of the device as $\omega_0 \approx \alpha \sqrt{GV/I} \approx \alpha \sqrt{G / \rho l_x l_y}$. $G$ is the shear modulus and $V = l_x l_y t_{\rm g}$ is the volume of the gyroscope. The proportionality factor $\alpha$ is on the order of 1/1000.

For a given gyroscope, controllable parameters that affect the signal is just $B_{x(y)}$, $t_{\rm f}$ and $i$. Striking feature is that the signal does not depend on the size of the device. This is because decrease in angular momentum is mostly compensated by the decrease in moment of inertia. One can then work with a microscopic sample of superfluid. This has an interesting implication in studying topological quantum phase transitions, a point we will return to later. With such a device, thin film of helium can be deposited on the planar surface of the gyroscope, instead of being held inside a container. Edges, in this case, are in contact with vacuum, not container walls. Thus specular edges are formed, an important criteria for having $L_{\rm edge} = N \hbar/2$ \cite{sau11,vol14}. The thickness $t_{\rm f}$ becomes the sole parameter to control the signal strength of the angular momentum.

\begin{figure}[t]
\centerline{\epsfxsize1.0\hsize\epsffile{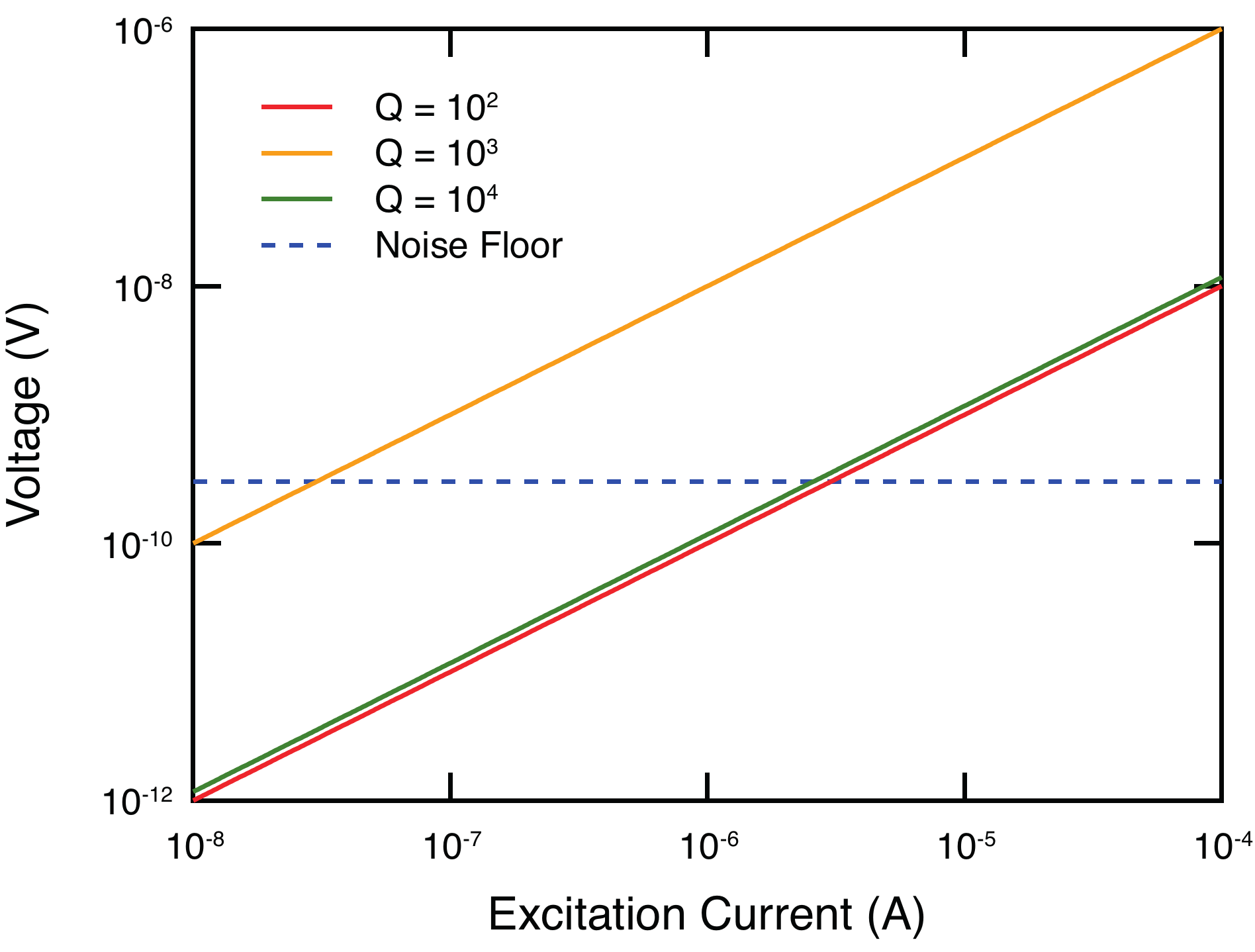}}
\label{sensitivity}
\caption {Estimated signal strength for three different values of $Q=Q_x = Q_y$ in comparison to the noise floor. Noise floor is dominated by the input noise of the cryogenic low noise amplifier$\rm{(\approx 300 pV/\sqrt{Hz})}$ where the measurement bandwidth is set to 1 Hz. $B = 300$ G and $t_f = 100$ nm are used in Eq.~(\ref{signal}). For a moderate quality factor of $10^3$ or less, the two orthogonal resonance frequencies are well-matched and the product $Q_x Q_y = 10^6$ enters into Eq.~(\ref{signal}). When the quality factor is as high as $10^4$, due to the narrow resonances, matching two frequencies becomes more difficult and only $Q_y =10^4$ is relevant. This is why the signal for $Q=10^4$ is estimated to be smaller than that for $Q=10^3$.}
\end{figure}

For a Si$_3$N$_4$ membrane of thickness 100 nm for gyroscope, mass density $\rho$ is roughly 3 g/cm$^3$ and shear modulus $G$ is on the order of 100 GPa. The signal to be detected is estimated to be
\begin{equation}
\label{signal}
V \approx \left( \frac{B}{300~\mathrm G}\right)^2 \left( \frac{i}{1~\mathrm A} \right) \left( \frac{t_{\rm f}}{1~\mu \mathrm m} \right) Q_x Q_y  \times 10^{-7} ~\mathrm V.
\end{equation}
The product of the quality factors for the two oscillations serve as a gain, $G_{\rm dB} = 10 \log Q_x Q_y$, on the signal. If the two resonances cannot be matched, however, only the quality factor of the detection resonator $Q_y$ serves as the gain with $Q_x$ effectively becoming close to unity. When mechanical resonators are immersed in $^3$He-$A$, the nodal excitations is responsible for damping and quality factor is expected to range between 10$^2$ and 10$^4$ depending on the temperature and film thickness. With NbTi superconducting leads, critical current density is about 100 kA/cm$^2$. An electrical lead with 1 $\mu$m width and 100 nm thickness can carry up to 100 $\mu$A. A moderate $Q$ of 10$^2$ is expected to give large enough signal to measure the intrinsic angular momentum. (See Fig.~2 for more details.)

$^3$He will be deposited on the device utilizing three different techniques. At high temperatures of around 0.3 K, one can deposit thin film by evaporating the $^3$He bath \cite{cro90}. The technique is equivalent to metallic thin film deposition using an evaporator. It just happens to be effective at much lower temperature due to the very low boiling temperature of liquid $^3$He. Because any wetting of the fluid on to the gyroscope is, in effect, mass-loading on to a torsional oscillator, the film thickness can be accurately monitored by monitoring the frequency shift of the oscillation. Quality factor of Q = 10$^3$ translates into roughly 10 nm resolution in the thickness. With this method, controlling the film thickness from 10 nm to a few hundred nm becomes possible. Once this film is cooled below the superfluid transition temperature, $T_c$, one of the more peculiar properties of a superfluid can be used to control the thickness. Superfluid climbs up walls against gravity. If bulk liquid level is set inside a container, superfluid would creep up the wall and wet any surface it finds as shown in Fig.~3. By adjusting the bulk liquid level, the thickness of the superfluid film on the gyroscope can be controlled from about 30 nm to a few hundred nm \cite{sac85,sch98}. To deposit thicker films, voltage can be applied to interdigitated capacitors (IDCs) patterned on the face of the gyroscope. Dielectric pondermotive effect pulls the liquid onto the the gyroscope. Deposition of liquid affects the dielectric constant which can in turn used to measure the film thickness by measuring the capacitance. With IDCs, the film thickness grows up to a few micrometers \cite{sai03}.

\begin{figure}[b]
\centerline{\epsfxsize1.0\hsize\epsffile{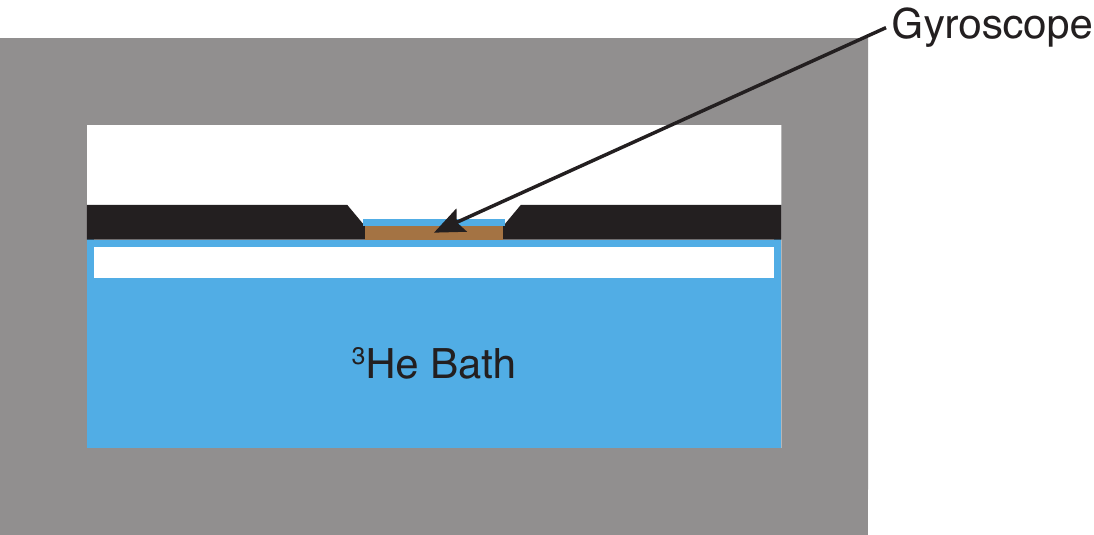}}
\label{film_deposit}
\begin{minipage}{1.0\hsize}
\caption {Gyroscope will be placed above $^3$He liquid bath as shown above. At temperature above 300 mK, evaporated $^3$He from the bath will be deposited on to the gyroscope surface. With $\sim$ 100 nm film, the device and the film has strong enough thermal link to the bath to follow the bath temperature. Once superfluidity is obtained, one can control the thickness of the film by adjusting the bath level. For thicker films, IDCs patterned on the gyroscope surface can be used to draw film as thick as a few $\mu$m.}
\end{minipage}
\end{figure}
\noindent

The controllability of the thickness has an important implication in studying the edge current. Firstly, as we have seen in Eq.~(\ref{magnetomotive}), the film thickness dictates the signal. One could test linear relation between the film thickness and the measured angular momentum by controlling the thickness in situ. In addition, the phase transition between $^3$He-$A$ and $^3$He-$B$ is driven by the thickness control. Because $^3$He-$A$ is stabilized through dimensional confinement, there is a critical thickness of about 1 $\mu$m below and above which $^3$He-$A$ and $^3$He-$B$ appears, respectively \cite{vor03}. The critical thickness is dictated by the coherence length scale at around $10 \xi \approx 1~\mu$m. As discussed earlier, $^3$He-$B$ is a time reversal protected helical TSC and the boundary carries spin current but not mass current. A transition between 2D D class (2D confined $^3$He-$A$) and 3D DIII class ($^3$He-$B$) TSCs would be marked by sudden appearance or disappearance of the angular momentum \cite{chi16}.

Because the signal strength does not depend on the size of the device, the effect of in-plane confinement can also be studied. A device can be fabricated with the dimension along one of the in-plane axis suppressed down to $\sim 10 \xi$. The order parameter suppression along the reduced dimension turns the superfluid into a quasi 1D polar phase. This marks a phase transition from 2D D class to 1D D class TSC \cite{chi16}. Hybridization of the edge states on the two edges that are brought close to each other opens up a gap and the edge current no longer flows. This is consistent with Majorana fermions localized on the two end points of the 1D superfluid. This transition must be accompanied by disappearance of the angular momentum as well. Controlling the dimension and geometry is potentially useful for engineering and observing interference of the edge current in the future.

For all of these proposed experiments to be successful, sample must be cooled down to sub-mK temperatures. Specific heat of Si$_3$N$_4$ is never measured below 80 mK \cite{lei98}. Nonetheless, the heat capacity of the gyroscope at 100 mK is at least two orders of magnitude smaller than that of $^3$He \cite{gre83} with $t_{\rm f} = 1~\mu$m. $^3$He, being an almost ideal Fermi liquid, has extremely high thermal conductivity at temperatures below 100 mK, as large as 10$^{-3}$ W/cm$\cdot$K \cite{gre84}. Given the specific heat of 100 mJ/K$\cdot$mol \cite{gre83}, cooling down from 100 mK to sub-mK can be done in $\sim 10$ seconds. Also, due to the large thermal conductance, relatively large amount of heat, $\sim$ 1 fW, injected into the gyroscope will not raise the temperature of the superfluid by more than 0.1 mK. The main source of heating would come from Joule heating of the wires and mechanical motion of the gyroscope dissipated into heat. For an ac driven current of 0.1 mA or less at around 10 kHz on a superconducting NbTi wire would suffer ac loss of about 100 aW \cite{wip68}. The kinetic energy of the gyroscope is dissipated into heat within the timescale given by the inverse of the resonance bandwidth. For a torsional motion with 0.1 mrad amplitude with 10 kHz resonant frequency, 10 aW of heat would be generated even for the quality factor as low as 10. In other words, the amount of heat released from the device is trivial and one should be able to keep superfluid $^3$He at sub-mK temperatures.

In summary, we propose a realistic experiment to measure the angular momentum rising from the edge current of a $p_x + ip_y$ TSC, based on a high sensitivity micro-electromechanical gyroscope. We believe this new technique gives a unique opportunity to measure the edge current of a topological superfluid and explore its properties, which will settle the long standing angular momentum paradox in $^3$He-$A$. The technique also holds an advantage over electronic TSCs. Majorana fermions are charge neutral and even if they are realized in a 2D electronic solid state systems, the edge current cannot be detected from the charge degree of freedom so easily \cite{bee16}. In comparison, a gyroscopic signal, a uniquely measurable quantity in superfluid $^3$He, gives direct evidence of the edge current \cite{sau11}. It will open up new avenues for TSC research as a testbed for dimensionality driven topological phase transitions and edge current interference.

We would like to thank J. A. Sauls, W. P. Halperin, Y. Lee and J. Pollanen for helpful discussions. This work is funded by Samsung Science \& Technology Foundation SSTF-BA1601-08.

\end{document}